%% file: vespa.tex
\documentclass{sig-alternate-05-2015}
\usepackage{graphicx}

\usepackage{epstopdf}

\usepackage{tabularx}

\usepackage{xparse}
\NewDocumentCommand{\ceil}{s O{} m}{%
  \IfBooleanTF{#1} 
    {\left\lceil#3\right\rceil} 
    {#2\lceil#3#2\rceil} 
}

\usepackage{algorithm,algpseudocode}

\usepackage[caption=true]{subfig}

\DeclareCaptionType{copyrightbox}

\makeatletter
\def\@copyrightspace{\relax}
\makeatother

\usepackage[nolist]{acronym}
\input{./acronyms.tex}

\begin{document}



\hyphenation{data-base pseu-do-nyms pseu-do-nym ano-ny-mi-ty a-no-ny-mi-za-tion e-quip-ped pse-udo-ny-mi-ty co-lo-gne in-fra-struc-ture Ker-be-ros}

\title{VeSPA: Vehicular Security and Privacy-preserving Architecture}

\numberofauthors{1}
\author{
\alignauthor Nikolaos Alexiou, Marcello Lagan\`a, Stylianos Gisdakis, \\
Mohammad Khodaei, and Panagiotis Papadimitratos \\ 
		\vspace{0.5em}
       \affaddr{Networked Systems Security Group}\\
       \affaddr{KTH Royal Institute of Technology}\\
       \affaddr{Stockholm, Sweden}\\
   		\vspace{0.5em}
       \email{\{alexiou, lagana, gisdakis, khodaei, papadim\}@kth.se}
}


\maketitle


\begin{abstract}

Standardization and harmonization efforts have reached a consensus towards using a special-purpose Vehicular Public-Key Infrastructure (VPKI) in upcoming Vehicular Communication (VC) systems. However, there are still several technical challenges with no conclusive answers; one such an important yet open challenge is the acquisition of short-term credentials, \emph{pseudonym}: how should each vehicle interact with the \acs{VPKI}, e.g., how frequently and for how long? Should each vehicle itself determine the pseudonym lifetime? Answering these questions is far from trivial. Each choice can affect both the user privacy and the system performance and possibly, as a result, its security. In this paper, we make a novel systematic effort to address this multifaceted question. We craft three generally applicable policies and experimentally evaluate the \acs{VPKI} system performance, leveraging two large-scale mobility datasets. We consider the most promising, in terms of efficiency, pseudonym acquisition policies; we find that within this class of policies, the most promising policy in terms of privacy protection can be supported with moderate overhead. Moreover, in all cases, this work is the first to provide tangible evidence that the state-of-the-art \acs{VPKI} can serve sizable areas or domain with modest computing resources. 



\end{abstract}

%
%
\printccsdesc



\input{intro}
\input{relatedwork}

\input{problemstatement}

\input{vpki}

\input{results}

\input{conclusion}

\bibliographystyle{abbrv}
\bibliography{biblio}  


\end{document}

%% file: acronyms.tex
\begin{acronym}
	\acro{C2C-CC}{Car-to-Car Communication Consortium}
	\acro{CRL}{Certificate Revocation List}
	\acro{ETSI}{European Telecommunications Standards Institute}
	\acro{VC}{Vehicular Communications}
	\acro{V2V}{Vehicle-to-Vehicle}
	\acro{V2I}{Vehicle-to-Infrastructure}
	\acro{PKI}{Public Key Infrastructure}
	\acro{VPKI}{Vehicular Public Key Infrastructure}
	\acro{RS}{Road-Side}
	\acro{PCA}{Pseudonym Certification Authority}
	\acro{LTCA}{Long-Term Certification Authority}
	\acro{RA}{Resolution Authority}
	\acro{CA}{Certification Authority}
	\acro{CRLs}{Certificate Revocation Lists}
	\acro{TRL}{Ticket Revocation List}
	\acro{RSI}{Road Side Infrastructure}
	\acro{AAA}{Authentication, Authorization and Accountability}
	\acro{WAVE}{Wireless Access in Vehicular Environments}
\end{acronym}

%% file: intro.tex
\section{Introduction}
\label{Sec:hotwisec-2013-intro}

\acp{VC} comprise vehicles and \ac{RSI} acting both as end-hosts and routers, interacting in ad-hoc manner using wireless communication technologies, such as 802.11 and cellular networks. Safe driving is the milestone application for \ac{VC}. Vehicles broadcast beacon messages in frequent time intervals to report on their location, velocity and other safety-critical information. Besides safety, proprietary applications like location-based services, tolling systems and leisure applications, are expected to be developed for \ac{VC}. Therefore, a mixture of service providers and mobile devices will interact with the \ac{VC}, essentially being part of it, and will therefore form the security and privacy challenges for vehicular networks. 

Message alternation and fabrication, as well as Denial of Service (DoS) pose important security challenges for \ac{VC} \cite{per}. Availability of the infrastructure through wireless communications is an additional network requirement for Vehicle-to-X communications that should operate under low response times. Additionally, \emph{Key Distribution} and \emph{Authentication} are important aspects for \ac{VC} that impose the existence of a \ac{CA} and eventually, secure hardware modules in the vehicles to manage the cryptographic keys \cite{preserve}. On the flip side of the coin,  authentication should be addressed with respect to vehicle \emph{Location Privacy} and \emph{Anonymity}, by protecting the vehicle from adversaries or trusted but curious infrastructure.

The current standards \cite{1609} and automotive industry directions \cite{c2c}, as well as research projects \cite{preserve}, address security and privacy challenges by suggesting an instantiation of a \ac{PKI}, known as \ac{VPKI}. Digital certificates signed by a trusted authority, allow the propagation of trust in the \ac{VPKI} hierarchy and also, enable anonymous mutual authentication between vehicles and the infrastructure. Short lived digital certificates, the pseudonyms, are adopted as the prevalent means to prevent the potential breach of vehicle privacy. However, anonymous authentication \textit{per se} cannot address the need for authorization and accountability posed by the large palette of future proprietary vehicular applications, and the current proposals should be enhanced towards this direction.  

In this work, we present the first implementation of a \ac{VPKI}, in order to secure \ac{VC} using a privacy-preserving architecture according to the standards. We present a \emph{kerberized} version of a \ac{VPKI} using cryptographic tickets to enable \ac{AAA} to the provided services. Our scheme offers credential management, while preserving the privacy against the \ac{VPKI} itself. Finally, we present an efficiency evaluation of our implementation and demonstrate its applicability. 

The remainder of this paper is organized as follows: in Sec.~\ref{Sec:hotwisec-2013-related} we present the related work, while in Sec.~\ref{Sec:hotwisec-2013-prblm} we define the problem statement. In Sec.~\ref{Sec:hotwisec-2013-vpki} we outline our architecture and protocols, while in Sec.~\ref{Sec:hotwisec-2013-Results} we demonstrate latency and efficiency results. We conclude the paper with a discussion and our future directions in Sec.~\ref{Sec:hotwisec-2013-conclusion}.

%% file: relatedwork.tex
\section{Related Work}
\label{Sec:hotwisec-2013-related}



Three anonymization schemas based on pseudonymous certificates and group signatures presented in~\cite{CalandPLH:C:2007}. A draft version of standards for secure \ac{VC} employing the pseudonym paradigm appeared in the IEEE 1609 family of standards for \ac{WAVE}~\cite{1609}.  Other standardization and harmonization efforts by the \ac{C2C-CC}~\cite{c2c} and \ac{ETSI}~\cite{ETSI} also converged towards the usage of pseudonymous certificates for privacy-preserving vehicular applications. The European Project SeVeCom~\cite{SeVeCom} defines the architecture for secure \ac{VC}. In addition, it addresses aspects such as key management and distribution, vehicle certification, and credential management.

The effectiveness of pseudonyms in preserving anonymity and location privacy for \ac{VC} is studied in~\cite{WiederMKP:C:2010, buttyan2007effectiveness}. Attackers with overwhelming monitoring capabilities can compromise privacy, but pseudonymous schemas undoubtedly offer improved resilience against adversaries. The impact of security on safety beaconing has been studied in~\cite{kargl:2008, PapadiCLH:C:08}. Although the current proposals for security and privacy rely on the implementation of a \ac{VPKI}, this is the first work to provide efficiency results and considers a \ac{AAA} solution.

Ticket-based authentication mechanisms, such as Kerberos~\cite{Kerberos}, centralize the identity management and accountability but do not offer anonymous service access. In~\cite{vtoken} a resolution approach using cryptographic tokens issued by a trusted authority is presented. However, the pseudonym acquisition protocol presented can compromise vehicle privacy (discussed later in Sec.~\ref{SubSec:hotwisec-2013-PRP}). In this work, we present a method of preserving the unlinkability of two consecutive requests and thus improving privacy. 



De-anonymization of the vehicles in case of user misbehavior is a requirement for safety applications in \ac{VC}~\cite{preserve,c2c}. Therefore, \ac{PKI} paradigms such as~\cite{Zhang:2005, Gu:2003} cannot be employed since they do not provide revocation or anonymity, respectively. Moreover, revocation schemas as presented in~\cite{Camenisch:2006,Tsang:2010} are not directly applicable in the \ac{VC} setting, since they do not offer identity resolution capabilities.

%% file: problemstatement.tex
\section{Problem Statement}
\label{Sec:hotwisec-2013-prblm}

Each vehicle is equipped with a tamper-resistant crypto-module able to perform advanced cryptographic operations, such as to digitally sign and encrypt messages. All packets transmitted by the vehicles should be authenticated. Packet authentication is not a guarantee of correctness, but the hardware security module greatly improves security as it reduces the chances of cryptographic keys being stolen. Each vehicle frequently broadcasts safety messages. 

We consider adversaries that deviate from the expected operation of the \ac{VC} protocols and can harm the security of the system and the privacy of its users in various ways. Launching impersonation attacks, the attacker claims to possess a legitimate identity and can fabricate messages or replay old packets. Attackers can deliberately change the content of packets to achieve erroneous or malicious behaviour. Such packet forgery attacks can result in serious implications for \ac{VC} especially when targeting safety beacons. Moreover, adversaries might try to gain access to \ac{VC} services, and eventually obtain valid credentials, for example pseudonyms. Non-repudiation is an important security property for \ac{VC}, especially for accountability purposes. Jamming in \ac{VC} is a \emph{low effort} attack that can be launched over small or wider geographical areas, but is out of the scope for this paper. Adversaries targeting vehicle privacy and anonymity by linking successive pseudonyms, can leverage on the information included in safety-beacons, in order to reconstruct real vehicles' whereabouts. For this, academia, industry, and standardization bodies have converged on the use of pseudonymous credentials for privacy protection. Moreover, privacy needs to be considered even in the presence of untrusted (i.e. honest-but-curious) infrastructure and misbehaving users. In the later case, the anonymity provided by the pseudonymous identifiers needs to be revoked.

All of the above underline the importance of secure and privacy-preserving credential management for safety applications in \ac{VC}. Nevertheless, given the near-deployment status of \ac{VC}, a whole ecosystem of non-safety services and applications is on the way. To facilitate their adoption by users, a \ac{VPKI} must offer them security (i.e. \ac{AAA} services) and protect the privacy of travellers/users against inference attacks and profiling. All these define the need for a scalable, modular and resilient \ac{VPKI} implementation whose services support, but can be extended beyond, the domain of safety-applications. This becomes critical given the absence of an implementation and evaluation of such an infrastructure. These points comprise the motivation and the scope of our work. We design, implement and evaluate a standard-compliant \ac{VPKI}, able to accommodate the security and privacy requirements for safety applications and to offer secure and privacy-preserving credential management to any other vehicular application.

%

%% file: vpki.tex
\section{The VPKI Architecture}
\label{Sec:hotwisec-2013-vpki}

In this section we present our architecture and the relevant protocols. We focus on the security and privacy aspects of our approach, and define a privacy-preserving pseudonym acquisition protocol which can be easily extended to support other vehicular services.

\subsection{Security \& Privacy Discussion}

Packets signed under the private key of the vehicle, residing inside the hardware security module, are then transmitted along with the corresponding certificate. The \ac{VPKI} architecture should support key management and certificate distribution, thus ensuring
(i) \ac{VC} message integrity,
(ii) message \& vehicle authentication in both \ac{V2I} and \ac{V2V}, and
(iii) non-repudiation of origin security properties.
Vehicles can establish secure channels (e.g., using TLS tunnels), thus achieving confidentiality against external eavesdroppers. Authorization and accountability is accomplished using \emph{tickets}; that is reusable proofs of access rights to a given service. Tickets are signed by a trusted authority to avoid forgery and integrity attacks as presented in Sec.~\ref{SubSec:hotwisec-2013-PRP}. We now discuss the usefulness of two types of certificates: 

\textbf{Pseudonyms.}
In order to preserve location privacy and anonymity in \ac{VC}, each vehicle possesses a set of short-lived pseudonyms, obtained by a trusted pseudonym provider. Each pseudonym has a lifetime ranging from seconds to hours, defined by the pseudonyms provider. A vehicle can decide to change the active pseudonym in order to prevent the tracking of its location. Safety beacons are digitally signed under the current pseudonym identity. By increasing the frequency of pseudonym changes, the chances for an adversary to launch a successful attack against privacy are reduced. 

\textbf{Long-term Certificates.}
A pseudonym acquisition protocol is necessary to obtain new sets of pseudonyms when the old ones are close to expire or have been already used. However for accountability and authorization purposes, the vehicle needs to be authenticated using its long-term identifier and then obtain anonymous authorization credentials, in the form of tickets. For this reason, each vehicle should be able to prove its real identity using a \emph{long-term identity}.


\subsection{Architecture Proposal}

Our scheme comprises the following three trusted \ac{CA}s, according to the terminology used in~\cite{preserve} and compatible with the definitions in~\cite{ETSI}:

\begin{itemize}
\item \textbf{\ac{LTCA}}:\\ The \ac{LTCA} is the issuer of the vehicle's long-term certificates and tickets.
\item \textbf{\ac{PCA}}:\\ The \ac{PCA} is the provider of the vehicle's pseudonyms.
\item \textbf{\ac{RA}}:\\ The \ac{RA} \emph{de-anonymizes} pseudonymous certificates in case of misbehavior detection.
\end{itemize}

The long-term certificate is a digital signature of the \ac{LTCA} over a set of vehicle-specific identifying data, a validity period $[t_s, t_e]$, and
the vehicle's long-term public key $K_v$: 
\begin{equation*}
	LT_v = Sig_{\text{LTCA}}(K_v, \text{data}_v, [t_s, t_e])
\end{equation*}
We assume that each vehicle $v$ has a long-term certificate $LT_v$ and the corresponding private key $k_v$ pre-installed in its hardware security module, as proposed in~\cite{PapadBHH:J:08}. The vehicle also obtains and stores a set of pseudonyms of the following form:
\begin{equation*}
	P^i_v = Sig_{\text{PCA}}(K^i_v, [t_s, t_e])
\end{equation*}
Pseudonyms also have a specified validity period $[t_s, t_e]$ and contain a public key $K^i_v$ for
verification. 

\subsection{Pseudonym Request Protocol}
\label{SubSec:hotwisec-2013-PRP}

We now describe the protocol for the vehicles to obtain pseudonyms from the \ac{PCA}. All communications are performed over a secure TLS tunnel, which guarantees confidentiality against external adversaries, and prevents tickets hijacking. For vehicle-to-\ac{PCA} communications one-way authentication of the server to the vehicle is used, in order to preserve the anonymity of the vehicle. In a nutshell, the protocol starts with the vehicle being authenticated by the \ac{LTCA} using its long-term credentials to obtain a \emph{ticket}. The ticket, $tkt$, does not contain any data attributable to the vehicle and it is of the form: 
\begin{equation*}
	tkt = Sig_{\text{LTCA}} ([{t_s}, {t_e}], \{S_1\}, \dots, \{ S_n \} ),
\end{equation*}
where $[{t_s}, {t_e}]$ is the ticket validity period and $S_i$ is a generic service identifier. By ensuring that $t_e$ does not exceed the subscription expiration time for any of the $S_i$ included in $tkt$, the \ac{LTCA} can guarantee that service subscription periods are not violated.

\begin{subequations}  \label{eq:hotwisec-2013-req_tkt}
\begin{align}
        V \longrightarrow LTCA:& \: Sig_{k_v}(t_1, \text{Request}) \: \| \: LT_v \\
        LTCA \longrightarrow V:& \: tkt
\end{align}
\end{subequations}

Initially, the vehicle issues a ticket request to the \ac{LTCA} in order to obtain access to the \ac{PCA}. The \ac{LTCA} checks the validity of the request, generates $tkt$ and sends it back to the vehicle. The vehicle then generates a set of private/public key pairs $(k^i_v, K^i_v)$ inside its hardware security module and sends the public keys $K^i_v$, along with $tkt$, to the \ac{PCA}.

\begin{subequations} \label{eq:hotwisec-2013-req_pse}
\begin{align}
	V \longrightarrow PCA&: t_3, tkt, \{ K^1_v, \dots, K^n_v \} \\
        PCA \longrightarrow V&: t_4, \{ P^1_v, \dots, P^n_v \}
\end{align}
\end{subequations}

The \ac{PCA} assesses the validity of the ticket and signs the received public keys $K^i_v$ using its private key. The pseudonyms $P^i_v$ are then sent back to the vehicle. The same ticket can be re-used for multiple pseudonym requests, or different service providers during its validity period. 

\textbf{Unlinkability of requests.}
We avoid signing pseudonym requests under the long-term or the current-pseudonym identities of the vehicle. In both cases the \ac{PCA} can breach vehicle privacy. In the first case, linking the issued pseudonyms to the long-term identifier is trivial; in the latter case, the \ac{PCA} is able to link the new set of issued pseudonyms with the one used for the request. Therefore the \ac{PCA} can link sets of pseudonyms and thus, compromise privacy. On the other hand, using a new \emph{ticket-per-request} can effectively protect vehicle privacy against the \ac{PCA}, since no linking is possible between the ticket, the long-term certificate, or any of the pseudonyms. Moreover, the vehicle can issue a request per pseudonym, thus restricting the ability of \ac{PCA} to link pseudonyms within a request. The proof of the unlinkability is straightforward and omitted here due to space limitations. 


\subsection{Pseudonym \& Token Revocation}

Pseudonyms and long-term certificates should be revoked in a number of different scenarios: for example when a vehicle is involved in an accident or misbehaves. Similarly, a ticket can be revoked to deny access to the service e.g., in case the ticket should not be reused. In order to keep the network \emph{up-to-date} in terms of the status of revoked certificates and tickets, \acp{CRL} are used. Revocation lists are publicly available, so that every entity in the \ac{VC} network has access to them. \acp{CRL} are digitally signed with the private key of the authority that issues them. The \ac{PCA} signs the revocation lists containing the revoked pseudonyms and the \ac{LTCA} the \acp{CRL} containing the long-term certificates. The dissemination of the \acp{CRL} is orthogonal to this work. Equivalently, \ac{TRL} can be used for ticket revocation, published by the \ac{LTCA} in case of ticket revocation. We omit further discussions on ticket and certificate revocation in this work because of the limited space. 

\subsection{Resolution Protocol}
\label{SubSec:hotwisec-2013-Revocation}

Due to the safety critical nature of \ac{VC}, the revocation of anonymous credentials is not sufficient \textit{per se} and complete vehicle de-anonymization is required. The resolution protocol is executed with the \ac{RA} acting as a coordinator between the \ac{PCA} and the \ac{LTCA}. The \ac{PCA} reveals to the \ac{RA} the link between the pseudonyms and the anonymous ticket. Then, the \ac{LTCA} reveals the link between the the ticket the vehicle's real identity. Therefore, the \ac{RA} can combine both pieces of information and perform the resolution. 

The \ac{RA} generates a digitally signed \textit{resolution request} to the \ac{PCA}. The request includes the pseudonym $P^i_v$ (or the set of pseudonyms) that have to be resolved. The \ac{PCA} retrieves all the pseudonyms that were issued as a result of the same vehicle pseudonym acquisition request from its database, along with the corresponding ticket $tkt$.  

\begin{subequations} \label{eq:hotwisec-2013-res}
\begin{align}      
        RA \longrightarrow PCA : \: &Sig_{\text{RA}}(P^i_v, t_1) \\
        PCA \longrightarrow RA : \: &Sig_{\text{PCA}}(tkt, t_2)
\end{align}
Having received the ticket $tkt$ the \ac{RA} forwards it to the \ac{LTCA}, which can in turn reveal the corresponding long-term identity of the vehicle. Mappings between issued tickets and the corresponding long-term identifiers exist in the database of the \ac{LTCA}.
\begin{align}      
	RA \longrightarrow LTCA : \: &Sig_{\text{RA}}(tkt, t_3) \\
	LTCA \longrightarrow RA : \: &Sig_{\text{LTCA}}(LT_v, t_4)
\end{align}
\end{subequations}

With the completion of the protocol, the long term identity $LT_v$ is resolved and the vehicle's pseudonyms have been revoked. Revocation is performed according to the previous section, which will eventually evict the vehicle from the \ac{VC} network. The \ac{LTCA} should also invalidate the received tickets by including them in the \ac{TRL}, to prevent adversaries from distributing tickets among each-other.
\\

%% file: results.tex
\section{Results}
\label{Sec:hotwisec-2013-Results}

\begin{figure}[!t]
\centering
\includegraphics[width=1\columnwidth]{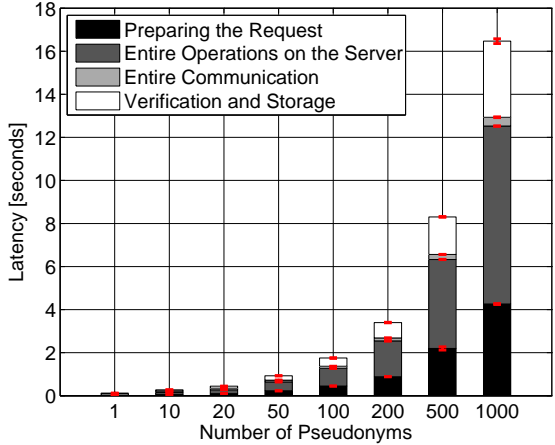}
\caption{Latency to obtain pseudonyms in seconds (per vehicle).}
\label{fig:hotwisec-2013-first}
\end{figure}

In this section we present the performance of the proposed \ac{VPKI} architecture. CAs were implemented using OpenCA, on separate servers equipped with an Intel Xeon Dual-Core $3.4$ GHz processor and $8$ Gbytes of RAM. All \ac{V2I} and Infrastructure to Infrastructure links are secured with TLS, while the study of the communication channels are out of the scope of this paper. ECC-256 keys are used for both infrastructure and vehicle certificates. Our implementation is compatible with the IEEE 1609.2 draft proposal~\cite{1609}. The ticket size is $498$ bytes and the pseudonym size is $2.1$ KBytes.

\textbf{Vehicle: Pseudonym Request.}
In Fig.~\ref{fig:hotwisec-2013-first}, we present latency results for acquiring a set of pseudonyms from the \ac{PCA}. The vehicle needs $73,4$ ms to obtain a new ticket from the \ac{LTCA} (eq.~\ref{eq:hotwisec-2013-req_tkt}). To acquire one pseudonym the vehicle needs $120$ ms and $3,400$ ms for $200$ pseudonyms (eq.~\ref{eq:hotwisec-2013-req_pse}). For requests of $1,000$ pseudonyms, which should be sufficient for a relatively long period or time (e.g., for a day if the pseudonym lifetime is around 1 minute), we observe that the total latency is $16,460$ ms. $50\%$ of the total latency concerns \ac{PCA} side operations, and $26\%$ is devoted on the preparation of the query, for examples the creation of private/public keys and digital signatures over the public keys.

The preparation of the request can take place off-line, which can eventually reduce the total time by $4,260$ ms (darkest bar in Fig.~\ref{fig:hotwisec-2013-first}).
Excluding the verification and storage time that occurs at the vehicle, the total processing time (communication plus operation on the server) to obtain $1,000$ pseudonyms is reduced to $8,670$ ms. Results suggest that our approach is efficient. Additionally, taking into consideration the fact that the vehicles will be equipped with hardware accelerators~\cite{preserve}, we can conclude that the time required for a vehicle to obtain a pseudonym will be significantly reduced.

\begin{table}[!h]
\scriptsize 
\begin{center}
\begin{tabularx}{\columnwidth}{|p{2.5cm}|>{\centering} X|>{\centering}X|X|X|X|X|} \hline
\textbf{Pseudonyms Req.} & \textbf{1} & \textbf{100} & \textbf{1.000} & \textbf{5.000} & \textbf{20.000} \\ \hline
\textbf{Signature Ver.} & $0,004$ & $0,361$ & $3,3618$ & $18,09$ & $72,33$ \\ \hline
\textbf{Pseudonyms Gen.} & $0,004$ & $0,349$ & $3,34$ & $17,72$ & $70,9$ \\ \hline
\textbf{Total Time} & $0,02$ & $0,817$ & $8,826$ & $41,672$ & $167,3$ \\ \hline
\end{tabularx}
\caption{Latency to issue pseudonyms in seconds by the PCA}
\label{Tbl:hotwisec-2013-pcalat}
\end{center}
\vspace{-5ex}
\normalsize
\end{table}

\textbf{PCA: Pseudonym Issuance.}
Table ~\ref{Tbl:hotwisec-2013-pcalat} shows the time needed by the \ac{PCA} to process pseudonym requests from vehicles. The processing time includes the verification of the received request (including ticket verification), pseudonym generation time and other relevant \ac{PCA} operations (e.g., storage and handling of the received public keys). For a total of $5,000$ pseudonym requests issued by multiple vehicles, $41,672$ ms are needed. For $20,000$ pseudonyms the server needs $167,300$ ms. It is straightforward that the pseudonym's lifetime is a determinant factor for the \ac{PCA}'s workload.

\begin{figure}[!t]
\centering
\includegraphics[width=1\columnwidth]{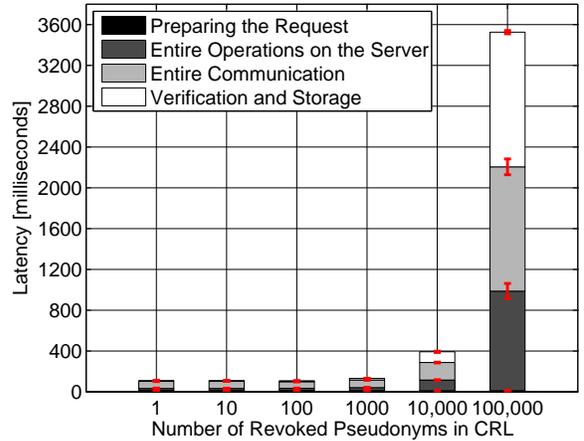}
\caption{Latency to obtain \acp{CRL} (per vehicle).}
\label{fig:hotwisec-2013-sec}
\end{figure}

\textbf{CRL Distribution.}
Fig.~\ref{fig:hotwisec-2013-sec} shows the time needed by a vehicle to obtain the \acp{CRL} of revoked pseudonyms. The preparation of the request by a vehicle takes $11$ msecs. The \textit{Server Operations} time corresponds to the generation of the CRL (including signing it) at the \ac{PCA}. We observe that latency increases with the number of entries in the CRL. For large chunks of information (e.g., $100,000$ entries in the CRL) the communication time is an important fraction of the total time; $1,218$ ms for $100,000$ entries in the CRL. For the latter case, the verification of the \ac{PCA}'s signature and the storage of the obtained CRL, can take up to $1,324$ ms.

\begin{table}[!h]
\scriptsize 
\begin{center}
\begin{tabularx}{\columnwidth}{|p{3.8cm}|>{\centering}X|>{\centering}X|>{\centering}X|>{\centering}X|X|>{\centering}X|} \hline
\textbf{Pseudonyms Resolved} & \textbf{1} & \textbf{10} & \textbf{50} & \textbf{100} & \textbf{200} \\ \hline
\textbf{Pseudonyms Prov. (PCA)} & $73$ & $135$ & $304$ & $516$ & $922$ \\ \hline
\textbf{Identity Prov. (LTCA)} & $9$ & $10$ & $15$ & $20$ & $57$ \\ \hline
\textbf{Resolution Auth. (PRA)} & $265$ & $348$ & $604$ & $916$ & $1598$ \\ \hline
\end{tabularx}
\caption{Resolution latencies in milliseconds; PCA, LTCA \& PRA}
\label{Tbl:hotwisec-2013-rsl}
\end{center}
\vspace{-5ex}
\normalsize
\end{table}

\textbf{Certificate Resolution.}
Certificate resolution (eq.~\ref{eq:hotwisec-2013-res}) times are presented in Table \ref{Tbl:hotwisec-2013-rsl}. Calculation times include server side operations (e.g., fetching the requested certificate from the database), sign and publish the certification list. The \ac{LTCA} has the lowest overhead, since the number of tickets is less than the number of pseudonyms that need to be retrieved from the databases of the \ac{LTCA} and \ac{PCA} respectively. The resolution of $200$ pseudonyms takes less than $1,000$ ms for the the \ac{PCA}, and we believe that our resolution protocol does not introduce a significant overhead for the \ac{VPKI}. The \ac{RA} has the highest workload during the resolution process ranging from $265$ ms (for $1$ pseudonym) to $1,598$ ms (for $200$ pseudonyms).

%% file: conclusion.tex
\section{Conclusion}
\label{Sec:hotwisec-2013-conclusion}

In this paper, we presented the implementation of a distributed \ac{VPKI} architecture in order to provide security and privacy protection in \ac{VC}. We proposed the use of tickets to guarantee unlinkability between consecutive vehicle requests for pseudonyms, when a new ticket is used for each request. To the best of our knowledge, this is the first work that provides \ac{AAA} capabilities for a \ac{VPKI} according to the current standards and the privacy requirements. Part of our future work includes the integration of relevant privacy-preserving methods and anonymous authentication techniques in our protocols. We believe that our scheme is efficient, applicable and thus, it can pave the road towards secure and privacy-preserving \ac{VC}.

%% file: vespa.bbl
\begin{thebibliography}{10}

\bibitem{buttyan2007effectiveness}
L.~Butty{\'a}n, T.~Holczer, and I.~Vajda.
\newblock {O}n the {E}ffectiveness of {C}hanging {P}seudonyms to {P}rovide
  {L}ocation {P}rivacy in {VANET}s.
\newblock In {\em European Workshop on Security in Ad-hoc and Sensor Networks},
  pages 129\textendash141, July 2007.

\bibitem{CalandPLH:C:2007}
G.~Calandriello, P.~Papadimitratos, J.-P. Hubaux, and A.~Lioy.
\newblock {E}fficient and {R}obust {P}seudonymous {A}uthentication in {VANET}.
\newblock In {\em Proceedings of the {ACM} {I}nternational {W}orkshop on
  {V}ehicular {A}d hoc {N}etworks ({VANET})}, pages 19\textendash28, Sep.
  2007.

\bibitem{Camenisch:2006}
J.~Camenisch, S.~Hohenberger, M.~Kohlweiss, A.~Lysyanskaya, and M.~Meyerovich.
\newblock {H}ow to {W}in the {C}lonewars: {E}fficient {P}eriodic n-{T}imes
  {A}nonymous {A}uthentication.
\newblock In {\em ACM CCS}, pages 201\textendash210, Oct. 2006.

\bibitem{c2c}
{C}ar-to-{C}ar {C}ommunication~{C}onsortium {(C2C-CC)}, Jan. 2013.

\bibitem{ETSI}
{ETSI TR 102 638}.
\newblock {I}ntelligent {T}ransport {S}ystems ({ITS}); {V}ehicular
  {C}ommunications; {B}asic {S}et of {A}pplications; {D}efinitions, June 2009.

\bibitem{Gu:2003}
J.~Gu, S.~Park, O.~Song, J.~Lee, J.~Nah, and S.~Sohn.
\newblock {M}obile {PKI}: {A} {PKI}-{B}ased {A}uthentication {F}ramework for
  the {N}ext {G}eneration {M}obile {C}ommunications.
\newblock In {\em Australasian Conference on Information Security and Privacy},
  pages 180\textendash191, July 2003.

\bibitem{1609}
{IEEE 1609.2}.
\newblock {D}raft {S}tandard for {W}ireless {A}ccess in {V}ehicular
  {E}nvironments - {S}ecurity {S}ervices for {A}pplications and {M}anagement
  {M}essages, Jan. 2012.

\bibitem{kargl:2008}
F.~Kargl, E.~Schoch, B.~Wiedersheim, and T.~Leinm{\"u}ller.
\newblock {S}ecure and {E}fficient {B}eaconing for {V}ehicular {N}etworks.
\newblock In {\em {P}roceedings of the 5th ACM International Workshop on
  Vehicular Ad Hoc Networks}, pages 82\textendash83, Sep. 2008.
  USA.

\bibitem{SeVeCom}
A.~Kung.
\newblock {S}ecurity {A}rchitecture and {M}echanisms for {V2V/V2I},
  {S}e{V}e{C}om - {D}eliverable 2.1, Feb. 2008.

\bibitem{Kerberos}
B.~Neuman and T.~Ts'o.
\newblock {K}erberos: {A}n {A}uthentication {S}ervice for {C}omputer
  {N}etworks.
\newblock {\em IEEE Communications Magazine}, 32(9):33\textendash38, Sep.
  1994.

\bibitem{PapadBHH:J:08}
P.~Papadimitratos, L.~Buttyan, T.~Holczer, E.~Schoch, J.~Freudiger, M.~Raya,
  Z.~Ma, F.~Kargl, A.~Kung, and J.-P. Hubaux.
\newblock {S}ecure {V}ehicular {C}ommunication {S}ystems: {D}esign and
  {A}rchitecture.
\newblock {\em IEEE Communications Magazine}, 46(11):100\textendash109, Nov.
  2008.

\bibitem{PapadiCLH:C:08}
P.~Papadimitratos, G.~Calandriello, J.-P. Hubaux, and A.~Lioy.
\newblock {I}mpact of {V}ehicular {C}ommunications {S}ecurity on
  {T}ransportation {S}afety.
\newblock In {\em IEEE INFOCOM Workshops}, pages 1\textendash6, Apr. 2008.
  Phoenix, AZ.

\bibitem{per}
B.~Parno and A.~Perrig.
\newblock {C}hallenges in {S}ecuring {V}ehicular {N}etworks.
\newblock In {\em Proceedings of Workshop on Hot Topics in Networks
  (HotNets-IV)}, Nov. 2005.

\bibitem{vtoken}
F.~Schaub, F.~Kargl, Z.~Ma, and M.~Weber.
\newblock {V}-tokens for {C}onditional {P}seudonymity in {VANET}s.
\newblock In {\em IEEE WCNC}, NJ, USA, Apr. 2010.

\bibitem{preserve}
J.~P. Stotz, N.~Bi{\ss}meyer, F.~Kargl, S.~Dietzel, P.~Papadimitratos, and
  C.~Schleiffer.
\newblock {S}ecurity {R}equirements of {V}ehicle {S}ecurity {A}rchitecture,
  {PRESERVE} - {D}eliverable 1.1, June 2011.

\bibitem{Tsang:2010}
P.~P. Tsang, M.~H. Au, A.~Kapadia, and S.~W. Smith.
\newblock {BLAC}: {R}evoking {R}epeatedly {M}isbehaving {A}nonymous {U}sers
  {W}ithout {R}elying on {TTP}s.
\newblock {\em ACM Transactions on Information and System Security (TISSEC)},
  13(4):39, Dec.

\bibitem{WiederMKP:C:2010}
B.~Wiedersheim, Z.~Ma, F.~Kargl, and P.~Papadimitratos.
\newblock {P}rivacy in {I}nter-{V}ehicular {N}etworks: {W}hy {S}imple
  {P}seudonym {C}hange is not {E}nough.
\newblock In {\em IEEE International Conference on Wireless On-demand Network
  Systems and Services}, pages 176\textendash183, Feb. 2010.
  Slovenia.

\bibitem{Zhang:2005}
Y.~Zhang, W.~Liu, W.~Lou, Y.~Fang, and Y.~Kwon.
\newblock {AC-PKI}: {A}nonymous and {C}ertificateless {P}ublic-{K}ey
  {I}nfrastructure for {M}obile {A}d {H}oc {N}etworks.
\newblock In {\em IEEE International Conference on Communications (ICC)},
  volume~5, pages 3515\textendash3519, May 2005.

\end{thebibliography}
